\documentstyle[11pt,paspconf,epsf]{article}
\newcommand{\HI}{H\thinspace\protect\footnotesize I\protect\normalsize}

\newcommand{\B}{{$B$}}

\newcommand{\II}{{$I_c$}}
\newcommand{\J}{{$J$}}
\newcommand{\K}{{$K_s$}}
\newcommand{\tfr}{Tully\,--\,Fisher relation}

\newcommand{\kms}{\,km\,s$^{-1}$}
\newcommand{\etal}{{\it et~al.}}
\newcommand{\cf}{{\it cf.\,}}
\newcommand{\eg}{{\it e.g.},\ }         
\newcommand{\msun}{\mbox{${\rm M}_\odot$}}

\def\la{\mathrel{\hbox{\rlap{\hbox{\lower4pt\hbox{$\sim$}}}\hbox{$<$}}}}
\def\ga{\mathrel{\hbox{\rlap{\hbox{\lower4pt\hbox{$\sim$}}}\hbox{$>$}}}}

\def\deg{{^\circ}}
\def\arcmin{\hbox{$^\prime$}}

\def\fm{\hbox{$.\!\!^{\rm m}$}}

\def\fdg{\hbox{$.\!\!^\circ$}}

\def\edcomment#1{\iffalse\marginpar{\raggedright\sl#1\/}\else\relax\fi}
\reversemarginpar

\begin{document}
\title{Multi-Wavelength Surveys for Galaxies Hidden by the Milky Way}
\author{Ren\'ee C. Kraan-Korteweg}
\affil{Departamento de Astronom\1a, Universidad de Guanajuato, 
Guanajuato GTO 36000, Mexico}

\begin{abstract}
The systematic mapping of obscured and optically invisible 
galaxies behind the Milky Way through complementary 
surveys are important in arriving at the whole-sky distribution of 
complete galaxy samples and therewith for our understanding of the 
dynamics in the local Universe. 
In this paper, a status report is given of the various deep optical, 
near infrared (NIR), and systematic blind \HI-surveys 
in the Zone of Avoidance, including a discussion
on the limitations and selection effects inherent to the different
multi-wavelength surveys and first results.

\end{abstract}


\section{Introduction}
Due to the foreground extinction of the Milky Way, galaxies
become increasingly fainter, smaller and are of lower surface brightness
as they approach the Galactic Equator. Although most of them are 
not intrinsically of low surface brightness, 
``whole-sky'' mapping of galaxies is required (a) in explaining 
the origin of the peculiar velocity of the
Local Group (LG) and the dipole in the Cosmic Microwave
Background,
(b) for our understanding of velocity flow fields such as
the Great Attractor in the Zone of Avoidance (ZOA) 
with a predicted mass excess of a few times $10^{16} \msun$ 
at ($\ell,b,v)\sim(320\deg,0\deg,4500$\kms, Kolatt \etal\ 1995), 
and (c) other suspected connections of nearby superclusters and
voids behind the Milky Way.

This not only concerns large-scale structures. Nearby massive 
galaxies behind the obscuration layer of the Milky Way could 
significantly change our understanding of the internal dynamics 
and mass derivations of the LG.

Dedicated searches for galaxies in about 25\% of the optically
obscured extragalactic sky so far revealed a number of important
features such as:

--  the nearby bright spiral galaxy Dwingeloo 1, a neighbor to
the LG (Kraan-Korteweg \etal\ 1994)

--  the Puppis cluster at ($\ell,b,v)\sim(245\deg,0\deg,$1500\kms) 
which may  contribute  at least 30\kms\ to the motion of the LG
{\it perpendicular} to the Supergalactic Plane (Lahav \etal\ 1993)

-- the massive Coma-like cluster A3627 
at ($\ell,b,v)\sim(325\deg,-7\deg,4800$\kms) which seems to 
constitute the previously unrecognized but predicted
density peak at the bottom of the potential well of the Great
Attractor (Kraan-Korteweg \etal\ 1996)

-- the 3C169 cluster at ($\ell,b,v)\sim(160\deg,0\deg,5500$\kms) 
connecting Perseus-Pisces and A569 across
the Galactic Plane (Chamaraux \etal\ 1990, Pantoja \etal\ 1997)

-- and the Ophiuchus (super-)cluster at
($\ell,b,v)\sim(0\deg,8\deg,8500$\kms) behind the Galactic Center 
(Wakamatsu \etal\ 1994).

In the following, I will review the current status of deep
optical searches behind the Milky Way, as well as the possibilities
and results given with the recent near infrared surveys and 
blind \HI-surveys.

\section{Optical Surveys}
Systematic optical galaxy catalogs are generally limited to the 
largest galaxies (typically with diameters 
D $\ga 1\arcmin$, \eg\ Lauberts 1982). These catalogs become,
however, increasingly incomplete as the dust thickens, 
creating a ``Zone of Avoidance'' in the distribution of galaxies
of roughly 25\% of the sky. Systematic deeper searches for partially 
obscured galaxies --  down to fainter magnitudes and smaller 
dimensions compared to existing catalogs -- were performed
with the aim to reduce this ZOA. These surveys are not biased 
with respect to any particular morphological type. 

The various survey regions are displayed in 
Fig.~1 (\cf\ Woudt 1998, for an extensive overview). Further
details and results on the uncovered galaxy distributions can be
found in
A:  Aquila and Sagittarius (Roman \etal\ 1996),
B:  Sagittarius/Galactic (Roman \& Saito 1997),
C:  Ophiuchus Supercluster (Wakamatsu \etal\ 1994),
D:  Galactic Center extension (Kraan-Korteweg, in progress),
E:  Crux and GA Region (Woudt \& Kraan-Korteweg 1999, in prep.),
F:  Hydra/Antlia Supercluster (Kraan-Korteweg 1999, in prep.),
G:  Hydra to Puppis Region (Salem \& Kraan-Korteweg, in progress)
H \& I :  Puppis (Saito \etal\ 1990, 1991),
I:  Perseus-Pisces Supercluster (Pantoja 1997),
J : northern crossing GP/SGP (Hau \etal\ 1996),
K:  northern ZOA (Seeberger \etal\ 1994).

\begin{figure}[h]
\hfil \epsfxsize 10cm \epsfbox{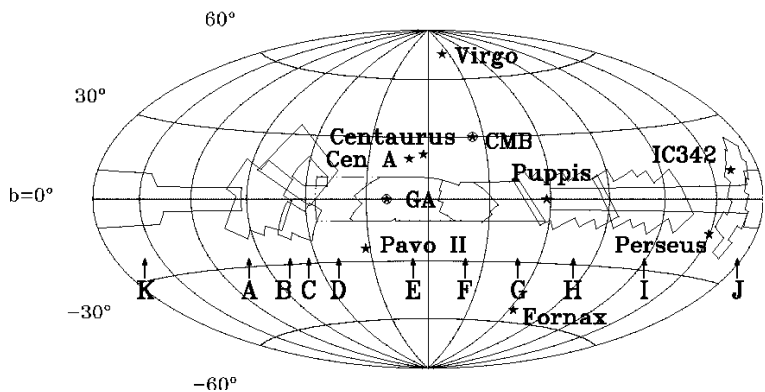} \hfil
\caption
{An overview of the different optical galaxy surveys in the ZOA centered
on $\ell=330\deg$. The labels identifying the search areas are explained 
in the text. Galaxy cluster positions (stars) and the CMB apex as well as
the core of the GA are marked.}
\end{figure}

Meanwhile, as Fig.~1 illustrates, nearly the whole ZOA has been surveyed 
systematically. With the mapping of over 50'000 previously unknown 
galaxies a considerable reduction of the ZOA was achieved.
Analysing the galaxy density as a function of the galaxy size, 
magnitude and/or morphology in combination with the foreground 
extinction has led to the identification of various important 
large-scale structures.
However, redshift follow-ups of well-defined samples are important
in tracing the large-scale structures in detail.
Such follow-up surveys have already revealed a number of 
dynamically important structures in the Zone of Avoidance
(\cf\ Introduction).
 
Although the various optical surveys are based on different plate
material and the criteria for inclusion in the respective surveys 
were not all identical, all surveys reveal the same
dependence on extinction: for extinctions in the blue of 
${\rm A}_{B} \ga 4-5^m$, the ZOA remains fully opaque (\cf\ top
panel of Fig.~3), leaving a strip of about $\pm 5\deg$ devoid
of galaxies.

\section{NIR-Surveys}
The extinction effects decrease with increasing wavelengths.
In the NIR passbands \II , \J\ and \K, the extinction compared 
to the blue is $A_{I_c} = 45$\%, $A_J\ = 21$\%, and $A_{K_s} =9$\%. 
Moreover, NIR surveys are sensitive to early-type galaxies -- 
tracers of massive groups and clusters missed 
in IRAS and \HI\ surveys --  and have little confusion with 
Galactic objects.
Here, the recent near infrared surveys, 2MASS (Skrutskie \etal\ 1997) 
and DENIS (Epchtein 1997) might provide new insight at low Galactic
latitudes. 

In unobscured regions, the density of galaxies per square degree 
for the completion limit of $B_J\le19\fm0$ 
is 110 (Gardner \etal\ 1996). However, the number counts in the 
blue decrease rapidly with increasing obscuration as $N(A_{\rm B}) 
\simeq 110 \times {\rm dex} (0.6\,[A_{\rm B}-19])\,$deg$^{-2}$.
In the NIR passbands \II , \J\ and \K\ of the DENIS survey, the 
counts for the respective completeness limits of
$I_{\rm lim}\!=\!16\fm0$, $J_{\rm lim}\!=\!14\fm0$, $K_{\rm lim}\!=\!12\fm2$
are considerably lower (30, 11, and 2, Mamon \etal\ 1997) but
-- as illustrated in Fig.~2 -- the decrease in number counts as a 
\begin{figure} [h]
\hfil \epsfxsize 8cm \epsfbox[20 161 564 532]{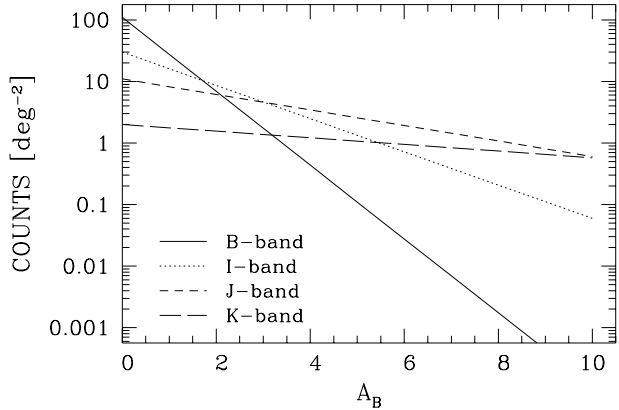}\hfil
\caption{Predicted galaxy counts in \B , \II , \J\ and \K\ as a function of
absorption in \B , for highly complete and reliable DENIS galaxy samples and
a $B_J \leq 
19^{\rm m}$ optical sample. }
\label{galctsplot}
\end{figure}
function of ``optical extinction'' is considerably slower.
The new cooling system for the focal instrument of DENIS installed
in 1997 led to an increase in the \K\ band counts of a factor of
two for a the completeness limit fainter by $\sim$ 0\fm5. 

Fig.~2 shows that the NIR becomes notably more efficient where the 
Milky Way becomes opaque in the optical ($A_B \ge 4-5^{\rm m}$, 
\cf\ previous section). 
At an extinction of $A_B \simeq 3-4^{\rm m}$, \J\ becomes superior 
to \II, while at $A_B \simeq 10^{\rm m}$, \K\ becomes superior to \J.
The cooled camera system will make the \K\ passband competitive 
with \J\ starting at $A_B \simeq 7^{\rm m}$. These are very
rough predictions and do not take into account any dependence on morphological
type, surface brightness, orientation and crowding, which may lower the
counts of actually detectable galaxies counts.

\subsection{First Results from DENIS}

To compare these predictions with real data,
Schr\"oder \etal\ (1997) and Kraan-Korteweg \etal\ (1998a)
examined the effeciency of uncovering galaxies at high 
extinctions with DENIS images. The results are
promising.

They found that down to intermediate latitudes and extinction
($|b| \ga < 5\deg$, $A_B \la 4-5^{\rm m}$),
optical surveys remain superior for identifying
galaxies.  However, the NIR luminosities and colors together with
extinction data from the NIR colors will prove invaluable in 
analysing the optical survey data and their distribution in redshift 
space, and in the final merging of these data with existing sky
surveys.  Despite the high extinction and the star crowding 
at these latitudes, \II , \J\ and \K\ photometry from the survey 
data can be successfully performed at these low latitudes and led,
for instance, to the preliminary $I_c^o$, $J^o$ and $K_s^o$ galaxy 
luminosity functions in A3627.

At low latitudes and high extinction
($|b| < 5\deg$ and $A_B \ga 4-5^{\rm m}$),
the search for `invisible' obscured galaxies on existing DENIS-images 
implicate that NIR-surveys can trace galaxies down to about $|b| \simeq
1\fdg5$. The \J\ band was found to be optimal for identifying galaxies up to
$A_B \simeq 7^{\rm m}$, although this might change in favour of \K\
with the new cooling system.  NIR surveys can hence further reduce the
width of the ZOA. Furthermore, this is the only tool that permits the 
mapping of early-type galaxies --- tracers of density peaks --- at 
high extinction.  

The analysis of DENIS images behind the ZOA is being pursued in a more
systematic way. Whether this will be performed by visual examination 
or whether galaxies can be successfully extracted using classical 
algorithms or artificial neural networks or a combination of both 
requires further exploration.

\section{Blind \HI\ surveys}
In the regions of the highest obscuration and infrared confusion,
the Galaxy is fully transparent to the 21-cm line radiation 
of neutral hydrogen. \HI-rich galaxies can readily be found 
at lowest latitudes through detection of their redshifted 
21-cm emission.
Only low-velocity extragalactic sources (blue- and redshifted) 
within the strong Galactic \HI\ emission will
be missed, and -- because of baseline ripple -- galaxies 
close to radio continuum sources.
Until recently, radio receivers were not sensitive and efficient 
enough to attempt large systematic surveys of the ZOA.
In a pilot survey with the late 300-ft telescope of Green Bank,
Kerr \& Henning (1987) surveyed $1.5\%$ of the ZOA and detected 16 
new spiral galaxies. 

\subsection{The Northern Zone of Avoidance}
Using the Dwingeloo 25m radio telescope, the whole northern Galactic
ZOA ($|b| \le 5\fdg25$) is being surveyed in the 21cm line for 
galaxies out to 4000~\kms (see also Rivers \etal, these proceedings). 
A shallow search (rms = 175~mJy) has been completed yielding five objects 
(Henning \etal\ 1998). This fast search for nearby massive galaxies
uncovered no major unknown Andromeda-like galaxy. The most exciting
discovery is the barred spiral galaxy Dwingeloo 1 
(Kraan-Korteweg \etal\ 1994), a new neighbour of the Local Group 
with one third of the Galaxy's mass.

The deeper survey (rms=40~mJy) is 60\% complete. 36 galaxies were
detected of which 23 were previously unknown, the most surprising
being the detection of a number of dwarfs at very low redshifts.
They lie close to the Sdm galaxy NGC 6946 at v=48~\kms , suggesting
a previously unrecognized nearby group or cloud of galaxies.

\subsection{The Southern Zone of Avoidance}
In March 1997, a systematic blind \HI\ survey began 
in the the southern Milky Way ($|b| \le 5\deg$) with the 
multibeam (MB) receiver (13 beams in the focal plane array) at the 
64\,m Parkes telescope. The survey covers the velocity range 
$-1200 \la v \la 12700$\kms\ and will have a sensitivity
of rms=6~mJy after Hanning smoothing. 

So far, a shallow survey based on 2 out of the foreseen 25 driftscan 
passages has been analysed (\cf\ Henning \etal\, these 
proceedings, and Kraan-Korteweg \etal\ 1998b). 107 galaxies were 
catalogued with peak \HI-flux densities of $\ga$80~Jy~\kms\ (rms=
15~mJy after Hanning smoothing). Though galaxies up to 6500\kms\ 
were identified, most of the galaxies (80\%) are quite local 
(v$<3500$\kms) due to the (yet) low sensitivity. 

Most detections are due to normal two-horned spiral galaxies. However,
ATCA follow-up observations of three very extended ($20\arcmin$
to $\ga 1\deg$), nearby (v $<$ 1500\kms) sources revealed them to
be interesting galaxies/complexes, with unprecedented 
low \HI\ column densities (\cf\ Staveley-Smith \etal\ 1998).

As in the northern \HI-survey, no Andromeda or other 
\HI-rich Circinus galaxy has been found lurking undetected 
behind the extinction layer of the southern Milky Way. 
Both \HI-surveys have, however, clearly proven the power of tracing 
spiral and \HI-rich dwarf galaxies through the deepest extinction
layer of the Milky Way (cf. Fig.~3 and 4, as well as Fig.~2 in
Rivers \etal, these proceedings).

\section{Conclusions}
Considerable progress has been made in mapping the galaxy distribution
with various multi-wavelength approaches. The continuing  surveys will
lead to a much more complete picture of the galaxy distribution in the 
``former'' ZOA. 

\begin{figure}[ht]
\hfil \epsfxsize 10cm \epsfbox[91 170 496 637]{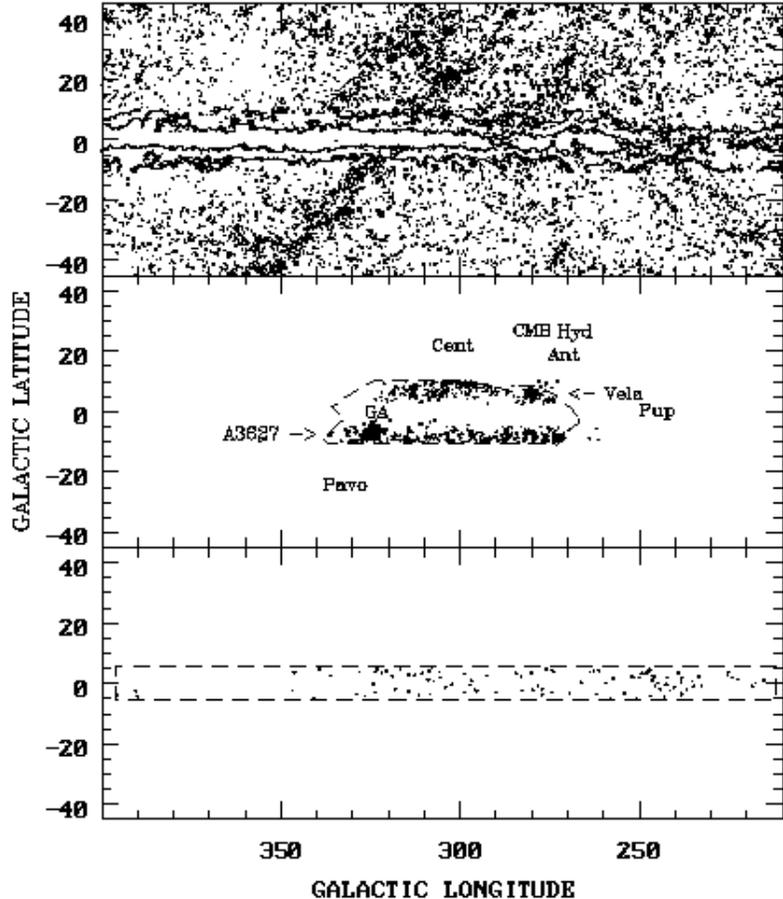} \hfil
\caption[]{Galaxies with v$<$10000 \kms.
Top panel: literature values (LEDA), superimposed
are extinction levels A$_B\sim1\fm5$  and $5^m$; middle panel:
follow-up redshifts (ESO, SAAO and Parkes) from deep optical
ZOA survey with locations of clusters and dynamically important
structures; bottom panel: redshifts from shallow MB-ZOA 
in \HI\ with the Parkes radio telescope. }
\end{figure}
How complementary the various multi wavelength approaches are is
illustrated in Fig.~3. The top panel
shows the distribution of all known galaxies with velocities 
v $\le 10000$\kms\ centered on the southern Milky Way. 
Although this constitutes an uncontrolled sample, it traces the main 
structures in the nearby Universe in a representative way. 
Note the near full lack of galaxy data for extinction levels
A$_B\sim1\fm5$ (outer contour).

The middle panel results from the follow-up observations of the 
optical galaxy search by Kraan-Korteweg and collaborators. Various
new overdensities could be unveiled at intermediate extinction levels
and low latitudes ($1\fm5 \la A_{\rm B} \la 5^m, 5\deg \la |b|
\la 10\deg$), but the innermost part of our Galaxy remains obscured 
(A$_B\ga4-5^m, |b|\la5\deg$). Here, the blind \HI\ data 
finally provide the missing link for large-scale structure studies 
as indicated with the results from the detections in the shallow 
survey of the Parkes MB survey (lower panel). 

\begin{figure}[h]
\hfil \epsfxsize 10cm \epsfbox[91 450 496 637]{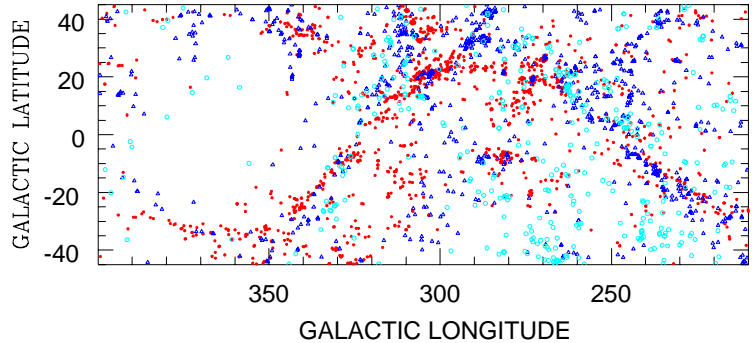} \hfil
\caption[]{Redshift slices from data in Fig.~3 for the
velocity range 500$<$v$<$3500 \kms. The open circles mark the 
nearest (500-1500), the triangles the medium (1500-2500) and
the filled dots the most distant (2500-3500) slice.}
\end{figure}
In Fig.~4, the data from the various surveys displayed in Fig.~3 are 
combined for the redshift range 500$<$v$<$3500~\kms. The upper
velocity limit 
reflects the depth achieved with the sensitivity of the shallow
MB \HI-survey.  For the first time, structures are 
visible  all the way across the ZOA. Note the continuity of the 
thin filamentary sine-wave-like structure that dominates the 
whole southern sky, and the prominence of the Local Void.  With 
the full sensitivity of the MB-survey, we will be able
to fill in the large-scale structures out to 10000~\kms.

The possiblities given for ZOA-research based on the currently
ongoing NIR are very promising -- and complementary in the sense
that they finally allow the uncovering of early-type galaxies
to low Galactic latitudes ($|b| \ga 1 - 1\fdg5$).
This is not the only addition NIR surveys provides. Schr\"oder \etal\ 
(1997) and Kraan-Korteweg \etal\ (1998a) have shown that
a fair fraction of heavily obscured spiral galaxies detected in blind 
\HI\ surveys can be reidentified on DENIS images (\cf, Fig.~5 in
Kraan-Korteweg \etal\ 1998a).
The combination of \HI\ data with NIR data allow the study of
the peculiar velocity field via the NIR \tfr\ ``in the ZOA'' 
compared to earlier interpolations of data adjacent to the ZOA
(Schr\"oder \etal, in progress).

A difficult task still awaiting us in the
future is the merging of ZOA data with catalogs outside the ZOA.
This will have to be done with care to obtain 'unbiased' whole-sky surveys.

\acknowledgements The collaborations with my colleagues in the various
multi-wavelength surveys --- P.A. Woudt with the
optical surveys, A. Schr\"oder and G.A. Mamon in the exploration of
the DENIS survey, W.B. Burton, P.A. Henning, O. Lahav and A. Rivers
in the northern ZOA \HI-survey (DOGS) and the HIPASS ZOA team members
R.D. Ekers, A.J. Green, R.F. Haynes, P.A. Henning, S. Juraszek,
M. J. Kesteven, B. Koribalski, R.M. Price, E. Sadler and 
L. Staveley-Smith in the southern ZOA survey --- are greatly appreciated.

\vfill
\end{document}